\documentclass[12pt,preprint]{aastex}

\usepackage{natbib}
\usepackage{graphicx}

\begin{document}

\title{Discovery of deuterated water in a young proto-planetary disk}

\author{ C. Ceccarelli\footnote{Laboratoire d'Astrophysique,
Observatoire de Grenoble - BP 53, F-38041 Grenoble cedex 09, France :
Cecilia.Ceccarelli@obs.ujf-grenoble.fr,
Bertrand.Lefloch@obs.ujf-grenoble.fr},
C. Dominik\footnote{Astronomical Inst. ``Anton Pannekoek'', Univ. of
Amsterdam, Kruislaan 403, NL-1098SJ Amsterdam :
dominik@science.uva.nl}, E. Caux\footnote{CESR CNRS-UPS, BP 4346,
31028 - Toulouse cedex 04, France : caux@cesr.fr}, B. Lefloch$^1$,
P. Caselli\footnote{INAF-Osservatorio Astrofisico di Arcetri, Largo
E. Fermi 5, I-50125 Firenze, Italy : caselli@arcetri.astro.it} }
\date{Received {\today} /Accepted {\today}}

\begin{abstract}
  We report the first detection of the ground transition of the
  deuterated water at 464 GHz in the young proto-planetary disk
  surrounding the solar type protostar DM Tau. The line is observed in
  absorption against the continuum from the cold dust in the disk
  midplane, with a line to continuum ratio close to unity. The
  observation implies that deuterated gaseous water is present, with a
  relatively large abundance ($\sim 3\times10^{-9}$), in the outer
  disk above the midplane, where the density is, within a factor
  ten, $\sim 10^6$ cm$^{-3}$ and the temperature is lower than about
  25 K. In these conditions, the H$_2$O condensation timescale is
  much smaller than the DM Tau disk age, and, therefore, water should
  be fully frozen onto the grain mantles. We suggest that UV photons
  and/or X-rays sublimate part of the mantles re-injecting the ices
  into the gas phase.  Even though there is currently no measurement
  of H$_2$O, we provide arguments that the HDO/H$_2$O ratio should be
  about 0.01 or larger, which would be hundreds of times larger than
  the values measured in Solar System objects.  This suggests the need
  of strong caution in comparing and linking the HDO/H$_2$O in Solar
  System and star forming environments.
 \end{abstract}

\def\fdep{\ensuremath{f_\mathrm{dep}}}
\def\kf{\ensuremath{k_\mathrm{f}}}

\keywords{Stars: formation -- Stars: protoplanetary disks -- Stars: Pre-main-sequence
-- ISM: molecules -- }


\section{Introduction}

Water is an important molecule to study during the formation of solar
type stars for several reasons.  It is one of the most
abundant molecules in the material surrounding a forming star, and it
is the most abundant molecule after H$_2$ and CO in the innermost
regions of the low mass protostellar envelopes
\citep{2000A&A...355.1129C}. Since water has a very large dipole
moment, it is a very efficient coolant of the gas, and can therefore
dominate its thermal balance \citep{1996ApJ...471..400C}. Finally,
because it can be a major reservoir of the oxygen element, water
also regulates the chemistry of other less abundant
oxygen-bearing molecules \citep{2002A&A...395..233R}.

Deuterated water in solar type forming stars is also important because
of the link with the comets and the Earth oceans.  Until recently, it
was thought that the terrestrial water came from the outer ($\geq 2.5$
AU) Solar System, brought by colliding bodies after the Earth
formation \citep{1995Icar..116..215O}.  This theory is now challenged
by a new class of dynamical simulations which theorize on the
simultaneous formation of the oceans and the Earth
\citep{2004Icar..168....1R, 2005Natur.435..466G}, but these
  ideas are still much debated.  A key indicator in the puzzle about
the ocean formation is provided by the so-called Standard Mean Ocean
Water (SMOW) HDO/H$_2$O ratio (Table \ref{tab:hdo}).  This value is
about ten times larger than the elemental Deuterium/Hydrogen ratio in
the Solar nebula, and similar to the values measured in comets and
carbonaceous chondrites (Table \ref{tab:hdo}).  The origin of the
HDO/H$_2$O ratio in both comets and meteorites is also very debated
\citep{1998Icar..133..147B}. It is often compared to what is observed
in the Interstellar Medium (ISM), either in the cold clouds from where
stars form or in the hot cores of {\it massive} protostars, which
possess HDO/H$_2$O values close to the SMOW
\citep{1998Sci...279..842M}.  However, recent observations in 
  protostars with {\it masses similar to the Sun} measure a HDO/H$_2$O
  value larger than the SMOW by about two orders of magnitude
  \citep[][see below]{2005A&A...431..547P}, so that the match is
  likely a coincidence.  Therefore, the question about the origin of
the HDO/H$_2$O ratio in comets, asteroids and the oceans is still
open.  In particular, it is not clear whether the HDO/H$_2$O is
preserved during the last phase before the formation of the planetary
system, the phase of the proto-planetary disk.  And no direct
observations exists so far of the water content and/or the HDO/H$_2$O
ratio in systems similar to the progenitor of the Solar System.

Molecular deuteration in the ISM and in star forming regions has
recently been the target of a flurry of activity, initiated by the
discovery of multiply deuterated molecules in solar-type protostars
\citep[e.g. see the review by][]{2002P&SS...50.1267C}, where D/H
ratios are enhanced by up to 13 orders of magnitude with respect to
the elemental D/H ratio \citep{2004A&A...416..159P}. This extreme
enrichment occurs during the initial stages of cloud collapse in the
Pre-Stellar-Phase \citep{2003ApJ...585L..55B}.  Low temperatures and
high densities lead to freeze-out of the molecules onto the grain
mantles, and to enhanced deuteration of H$_3^+$, the most abundant
molecular ion in such a gas, which ``transmits'' the deuteration to
the other molecules \citep{2003A&A...403L..37C}.  In the subsequent
protostar phase, heating of the dust leads to the sublimation of ices
in the inner $\sim$100 AU, injecting ice molecules back into the gas
phase \citep{2000A&A...355.1129C}.

Water, however, seems to follow a different path.  While the abundance
of singly deuterated isotopomers of H$_2$CO and CH$_3$OH are more
than 1/3 of the the main isotopomers, HDO is less than 3\% of H$_2$O
in solar-type protostars \citep{2005A&A...431..547P}.  Specifically,
HDO/H$_2$O is equal to 0.03 in the region where water ices sublimate,
and less than 0.002 in the outer envelope, where the water abundance
is dominated by gas-phase reactions.  The reason for this different
behavior most likely lies in the formation process.  Since water has a
large dipole, it freezes out onto the grain mantles already at $\sim
90$ K \citep{2001MNRAS.327.1165F}.  Therefore the reactions forming
water and water ice can occur at relatively high temperatures, when
the H$_3^+$ deuteration is limited.  On the other hand, formaldehyde
and methanol are believed to be formed by hydrogenation of frozen CO
\citep{1982A&A...114..245T}.  Since the condensation temperature of CO
is $\sim 25$ K \citep{2005ApJ...621L..33O}, this forces these
reactions to take place at very low temperatures. Under those
conditions, the enhanced D/H of the accreting atomic gas leads to high
deuterium fractionation of molecules formed by CO hydrogenation on
grains (e.g. Tielens \& Hagen 1982).

After the dispersion of the protostar envelope, a proto-planetary disk
is left over from which asteroids, comets and planets may form.  The
key questions to answer to improve our understanding of the origin of
the terrestrial water are: is there enough H$_2$O and HDO in a
  proto-planetary disk (similar to what could have been the Solar
  Nebula) to account for the water and ice present in the solar
  system?  What is the HDO/H$_2$O ratio across the disk?  Is the
HDO/H$_2$O of the pre-collapse and embedded protostar phases preserved
during the proto-planetary disk phase?  And, ultimately, what is the
origin of deuteration in comets, meteorites and oceans?

In this Letter we report the discovery of deuterated water in the disk
surrounding DM\,Tau, a T Tauri star at 140 pc from the Sun.  An
extended disk of molecular gas has been first discovered by
\cite{1994A&A...291L..23G}.  Subsequently \cite{1997A&A...317L..55D}
carried out a survey of different molecules.  The dust disk mass is $2
\times 10^{-4}$ M$_\odot$, and the age is estimated to be around 5
Myr.  Finally, the disk has a diameter of $12''$ (= 800 AU in radius).

\section{Observations and results}

\setcounter{footnote}{0}
The ground (1$_{0,1}$--\,0$_{0,0}$) transition of HDO at $\nu$ = 464.92452
GHz was observed on February 20, 22 and 28, 2005 with the JCMT near
the summit of Mauna Kea in Hawaii, USA. The observations were
performed with the dual-polarization W(C) receiver in single-sideband
mode. At the time of the observations, only one polarization was
active. It was connected to a unit of an autocorrelator providing a
bandwidth of 250 MHz for a spectral resolution of 156 kHz for some
scans, and a bandwidth of 500 MHz for a spectral resolution of 312 kHz
for others. All data were smoothed to the lowest spectral resolution,
yielding a velocity resolution of about 0.2 km s$^{-1}$. The
observations were performed in beam switching mode with a throw of
$180''$. Pointing and focus were regularly checked using planets or
strong sources, providing a pointing accuracy of about $3''$. The
telescope beam at 464 GHz is $11''$, and the main beam efficiency is
0.45 (as reported on the JCMT Manual:
http://www.jach.hawaii.edu/JCMT/spectral\_line/).

The data have been reduced with the GILDAS package CLASS. The total
integration time spent on the source was about 7.2 hours with a zenith
$\tau$ opacity at 225 GHz better than 0.06. Nevertheless, some scans
were taken at a high zenith angle, yielding to drift in calibration,
and higher noise. We therefore dropped the scans with observed $\tau$
$\ge$ 0.23 to build the final used dataset. The integration time of
this final dataset is 5.5 hours, and the main beam temperature rms is
equal to 33 mK in a 0.2 km s$^{-1}$ bin. 

\begin{figure}[htbp]
\centerline{\includegraphics[width=9.0cm]{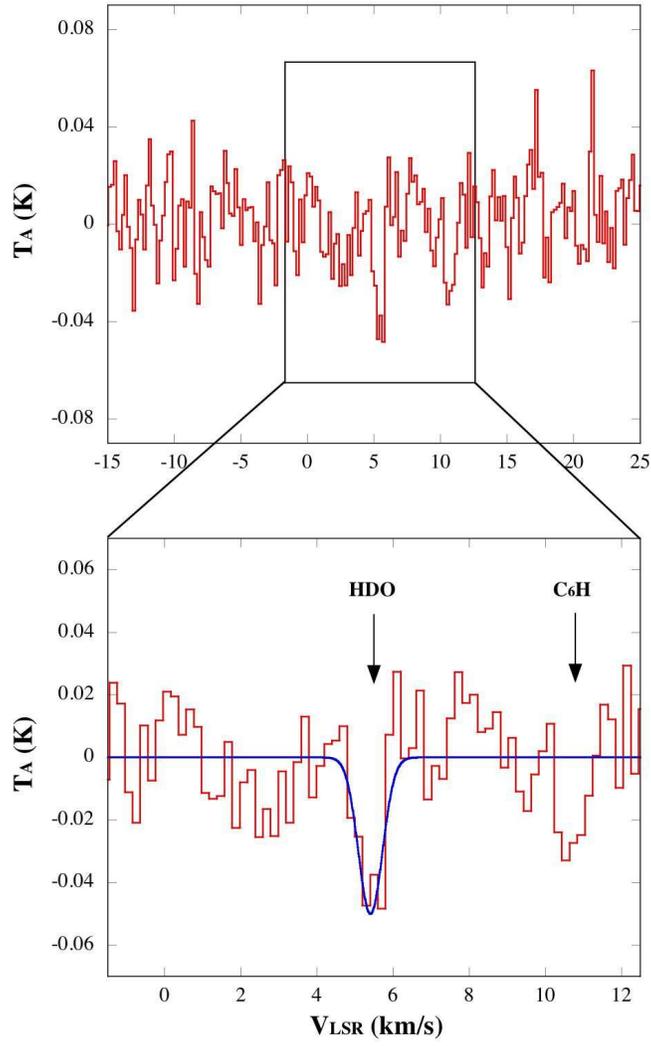}}
\caption{The observed line spectrum of DM Tau
at 464 GHz. The HDO ground transition, and the C$_6$H transition are
marked by lines.}
\label{fig:hdo} 
\end{figure}

Figure \ref{fig:hdo} shows the spectrum observed towards DM Tau. Two
absorption features are detected with a Signal-to-Noise (S/N) ratio
better than 3: the first one coincides exactly with the ground
transition of HDO (at 464.9245 GHz) when the systemic velocity (5.5 km
s$^{-1}$) of DM Tau is taken into account.  The other absorption
feature at 11 km s$^{-1}$ is due to a C$_6$H transition at 464.9172
GHz\footnote{The C$_6$H identification is firmly established thanks to
the presence of another absorption feature at 465.0511 GHz from the
same molecule in the spectrum, and to the detection of the same two
C$_6$H lines towards another disk source, during the same observation
run (Dominik et al. in preparation).}.  In order to check the robustness of
the detection, we analyzed all the obtained data in multiple subsets
of data, and the absorptions feature are present in all of them,
regardless of the definition of the datasets. This test ensures that
the features are indeed the result of summing all the spectra and not
due to a strong fluctuation in a few ones. All together, this gives
substantial support to the reality of the HDO absorption. We postpone
the discussion about the C$_6$H detection to a forthcoming article
(Dominik et al. in preparation). Here we focus on the HDO detection,
the first ever in a proto-planetary disk.  The velocity-integrated HDO
line is ($-0.075\pm0.019$) K km s$^{-1}$ (main beam temperature), and
the linewidth is (0.63$\pm$0.18) km s$^{-1}$.

A continuum at 464 GHz of $2$ Jy, derived by modeling of the Spectral
Energy Distribution (SED: see Figure \ref{fig:sed} and the text
below), gives a line to continuum ratio of 0.9 (i.e. the relative
depth of the absorption line is close to 1).  This corresponds to an
absorbing HDO column density equal to $8\times10^{12}$ cm$^{-2}$.
This number is based on the assumption that all HDO molecules are
in the ground state, which is a valid assumption for the involved gas
temperature and density.  The uncertainty associated with the derived
line to continuum ratio and HDO column density is around a factor 2,
when considering the uncertainty in the modeling and the observed line
absorption (close to unity).  However, the line may be optically
thick and, in which case the derived column density would be a lower
limit to the real HDO column density.

\section{Discussion}

In order to correctly interpret the meaning of the HDO line, the first
question to answer is: what is the location of the HDO molecules
causing the absorption?  Is it gas in the (cold) upper layers of the
outer disk (r$\geq$30 AU), or gas in the (warm) midplane close to the
star?  The answer to this question is straightforward. The photons
which are absorbed by the HDO molecules in the line of sight are
emitted by cold dust ($\sim10$ K), therefore by dust in the outer
midplane. As a consequence, {\it the HDO molecules must be in the gas
above the outer midplane}. Furthermore, because the line-to-continuum
ratio is close to unity, {\it the HDO gas must cover most of the
disk}.  Finally, the fact that the line is in absorption and not in
emission implies that the gas with HDO has a relatively low density,
below the critical density of the transition. The latter is
$\sim3\times10^9$ cm$^{-3}$ at 20 K \citep{2003A&A...408.1197G}.  {\it
We conclude that HDO (and therefore H$_2$O) vapor is present in the
outer disk above the midplane, at a density less than $\sim10^8$
cm$^{-3}$}.
Since only the gas towards us absorbs the 464
GHz photons, the total HDO molecules are twice the HDO column density
required to absorb the continuum, namely N(HDO)$\sim
1.6\times10^{13}$ cm$^{-2}$, to account for the other side of the
disk.  This implies a gaseous HDO mass in the disk equal to $\sim
2\times10^{23}$ gr, which is about 850 times the amount of HDO in the
terrestrial oceans.  Again, if the line is optically thick, all
numbers are lower limits.

\begin{figure}[htbp]
\centerline{\includegraphics[angle=90,width=9.0cm]{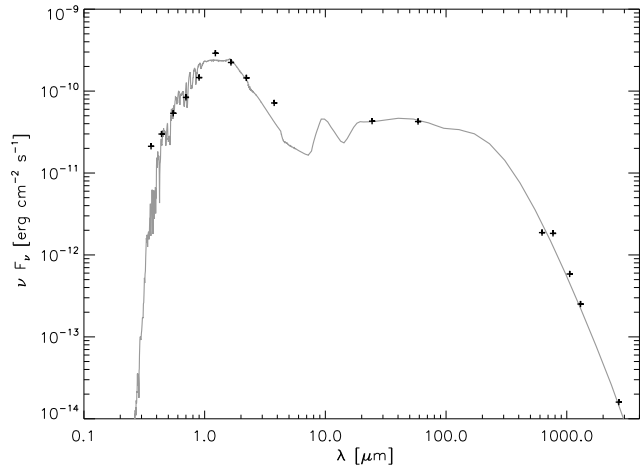}}
\caption{The observed Spectral Energy Distribution in DM Tau.  Crosses
  represent observations, while the solid line shows the modeled SED,
  using the model described in the text.  The disk has a mass of
  0.023 M$_{\odot}$, a surface density distribution $\propto r^{-1.5}$,
  in a disk reaching out to 800 AU from the star.  We used a stellar
  mass of 0.65 M$_{\odot}$, effective temperature of 3630 K and a
  luminosity of 0.28 L$_{\odot}$.  For fitting the observations, a
  distance of 140 pc has been assumed.  The structure resulting from
  this model is shown in Fig. \ref{fig:struct}.
}
\label{fig:sed}
\end{figure}

\begin{figure}[htbp]
\centerline{\includegraphics[angle=90,width=9.0cm]{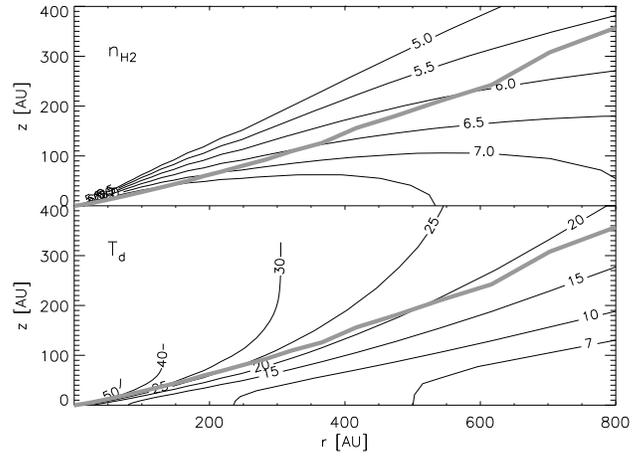}}
\caption{The logarithm of the density (upper panel) and the
temperature (lower panel) of the disk surrounding DM Tau, as derived by
modeling of the SED (Fig. \ref{fig:sed}). 
The thick line shows the disk surface, i.e. the location where the
disk reaches an optical depth of unity for grazing stellar photons.
}
\label{fig:struct} 
\end{figure}

Figure \ref{fig:struct} shows the physical structure of DM Tau,
derived by modeling the SED (Fig. \ref{fig:sed}). For that we used a
passive disk model with an inner hole \citep{2001ApJ...560..957D},
computed with 1+1D treatment of radiative transfer and self-consistent
vertical structure \citep{2002A&A...389..464D}. In ``standard''
conditions, H$_2$O molecules condense out onto the grain mantles at a
rate:
\begin{equation}\label{eq:1}
k_\mathrm{dep}=S \pi a_{\rm gr}^{2} n_\mathrm{g}
\left<v_\mathrm{H_2O}\right>
\end{equation}
and are released back into the gas phase by thermal evaporation, at a
rate (Hasegawa \& Herbst 1993):
\begin{equation}\label{kev}
k_\mathrm{ev}=\nu_{0} \exp[-E_\mathrm{b}/kT]
\end{equation}
In Eq. (\ref{eq:1}), $S$ is the sticking coefficient, $a_\mathrm{gr}$
is the mean grain radius (0.1 $\mu$m), $n_{\rm g}$ is the grain number
density (with respect to H$_2$, equal to $3.2\times10^{-12}$), and $
v_\mathrm{H_2O}$ is the H$_2$O thermal velocity.  In Eq. (\ref{kev}),
$\nu_0=10^{12}$s$^{-1}$ is the frequency of oscillation between
adsorbate and surface and $E_{\rm b}$ is the binding energy per
molecule.  Note that, in contrast to CO, cosmic rays do not contribute
significantly to the release of H$_2$O from the ice
\citep{1993MNRAS.261...83H}.  Taking the standard value for the
sticking coefficient (larger than 0.3), and the H$_2$O binding energy
measured by laboratory experiments \citep[$\sim 5600$
K;][]{2001MNRAS.327.1165F} leads to all the water frozen onto the
grain mantles at a distance larger of $\sim 30$ AU in much less than
the estimated age of DM Tau ($\sim 5$ Myr). This is a well known
effect in the literature of the Solar Nebula studies, giving rise to
the ``snow-line'' \citep{2005ApJ...620..994D}.  Therefore no vapor
water should be present in the outer disk.  However, {\it we do
observe HDO in the gas phase, in significant quantities and this is
the first important conclusion of the present work.}

In order to compute the HDO abundance, we need to estimate the H$_2$
column density in the region where the absorbing HDO is located.
Using the physical structure derived from the SED modeling
(Fig. \ref{fig:struct}), the H$_2$ column density of the gas at
600-800 AU is about $3\times10^{21}$ cm$^{-2}$, if the gas is located
at a height of about 300 AU (corresponding to a density of $\sim 10^6$
cm$^{-3}$). If the HDO absorption is located closer to the midplane,
at densities $\sim 10^7$ cm$^{-3}$, the H$_2$ column density would be
a factor three larger. In the unlikely case (see below) that the
absorption originates higher in the disk, the H$_2$ column density
could be a factor three lower. This implies a HDO abundance (with
respect to H$_2$) of $\sim 3\times10^{-9}$, with an uncertainty of
about a factor three, depending how deep in the disk the HDO gas lies.
Unfortunately, observations of H$_2$O are impossible from the ground,
and we need to wait for the advent of the Herschel Space Observatory
for an actual measure of the H$_2$O abundance in the gas
phase\footnote{Neither SWAS nor ODIN have the sensitivity enough for
the detection of water in DM Tau, unless the H$_2$O abundance is much
larger than $3\times10^{-7}$.}.  However, we can compare the derived
HDO abundance with the observations of H$_2$O in molecular clouds and
protostars, $\sim 3\times10^{-7}$
\citep{1997A&A...323L..25C,2003ApJ...582..830B,2002Ap&SS.281..139M},
and this would give HDO/H$_2$O equal to $\sim0.01$. If, on the other
hand, the water abundance is lower and more similar to what is
measured in cold molecular clouds, $\leq 10^{-8}$
\citep{2002ApJ...581L.105B}, the HDO/H$_2$O ratio would be larger than
0.3.  Therefore, the second conclusion of the present work is that
{\it the vapor HDO/H$_2$O ratio in the proto-planetary disk
surrounding DM Tau is probably larger than 0.01, unless the abundance
of the water is larger than $\sim 3\times10^{-7}$.}

In summary, the HDO detection proves that water vapor is present in
the outer disk, in the layers above the midplane where the
temperature is lower than about 25 K and density is, within a factor
ten, $\sim 10^6$ cm$^{-3}$, and that the HDO/H$_2$O ratio is likely
larger than 0.01. We explore now what this implies.  First, the
timescale for condensation of water molecules at $10^6$ cm$^{-3}$ is
$\sim 10^4$ yr (Eq. \ref{eq:1}) , whereas the estimated age of the
disk is of a few million years.  Therefore, the condensation of water
molecules onto grains is either slowed down significantly, or
countered by continuously ``sublimation'' of the icy grain mantles.
We are not aware of any physical process capable of slowing down
condensation of water onto the grain mantles.  We therefore consider
the possible mechanisms that would make the icy grain mantles
sublimate.  The first possibility is that UV photons photo-desorb
the icy mantles. Where UV photons penetrate, they also
photo-dissociate the water molecules, keeping the steady-state
abundance of water molecules low and limited to a narrow region. It is
unclear whether the resulting column density is enough to explain the
observations.  Shocks from possible turbulent mixing from the low to
the upper layers (vertical mixing) would be too slow ($\leq 1$ km
s$^{-1}$) and weak to give any appreciable effect. The same applies to
any accretion shock in the outer disk.  Another possibility is that
X-rays spot-heat the grain mantles making part of them sublimate
\citep{2001ApJ...561..880N}. Indeed, T Tau stars are known to be
strong X-rays emitters \citep{1999ARA&A..37..363F}, and a preliminary
analysis of CHANDRA observations suggests X-rays emission associated
with DM Tau (M. Guedel, private communication). Thus, X-rays would be,
in this sense, a natural explanation (see also
\cite{2004ApJ...614L.133B}), although it is difficult to quantify 
the amount of sublimated ice.

Finally, Table 1 lists the measurements of the HDO/H$_2$O ratio in
objects representing different stages of the formation of a solar type
star, compared to the measurements in the Solar System.  Although the
HDO/H$_2$O ratio in comets and carbon chondrites is 10 times larger
than the elemental D/H ratio, it is much smaller than the value
observed during the protostellar phase of a solar type star, and,
likely, also smaller than the value measured during the final phase,
namely the proto-planetary disk phase.  However, our observations
strictly refer to the outer disk, at a scale much larger than where
comets and carbon meteorits are thought to originate.  Therefore, 
it will be paramount to have the {\it spatial distribution} of the
HDO/H$_2$O ratio across the proto-planetary disks. Our observations
show that these studies can be done searching for the {\it absorption}
of the ground state of HDO and, likely, H$_2$O lines.  The future
interferometric instrument ALMA will represent a unique opportunity
for these studies.  As a final remark, the present study confirms that
caution is mandatory in linking the HDO/H$_2$O ratio in Solar System
objects to star forming environments.

\begin{acknowledgements}
  We wish to thank the JCMT staff for their precious help in doing the
  observations, and particularly J.Hough for his enthusiasm.  We
  acknowledge M.van den Ancker for a de-reddened SED of DM Tau,
  K.Dullemond for his help in constructing the disk model, and
  M.Guedel for the information about the X-rays emission of DM Tau.
  We also thank D.Bockelee-Morvan, D.Hollenbach, F.Selsis and R.Waters
  for very helpful comments on the manuscript.
\end{acknowledgements}

\clearpage


\clearpage
\begin{table}\label{tab:hdo}
 \begin{tabular}{|l|c|r|}
\hline
Source & HDO/H$_2$O & Reference\\ \hline
Solar Nebula & $1.5\times10^{-5}$ & 1 \\
Earth Oceans & $1.6\times10^{-4}$ & 2 \\
Carbonaceous Chondrites & $\sim1.5\times10^{-4}$ & 3 \\
Comets        & $\sim3\times10^{-4}$ & 4 \\
Proto-Planetary Disk & $\sim$0.01 & 5\\
Class 0 Protostar Sublimated Ices & $\sim 0.03$ & 6\\
\hline
 \end{tabular}
 \caption{The HDO/H$_2$O ratio in sources representing different
   stages of the formation of a solar type star, together with the
   values measured in the Solar System. References: 1-
   \cite{geiss1993}; 2- \cite{dewiietal1980}; 3-
   \cite{2000SSRv...92..201R}; 4- \cite{1998Sci...279..842M} ;5-
   present work; 6- \cite{2005A&A...431..547P}.  }
\end{table}

\clearpage

\clearpage

\clearpage

\end{document}